\documentclass[journal=jacsat,manuscript=article,layout=twocolumn]{achemso}

\usepackage{color}
\usepackage[dvipsnames]{xcolor}
\usepackage{placeins}
\usepackage{float}
\usepackage{multirow}
\usepackage{amsmath}

\author{E. Carlon}
\affiliation{KU Leuven, Institute for Theoretical Physics, Celestijnenlaan
200D, 3001 Leuven, Belgium}
\email{enrico.carlon@kuleuven.be}

\author{H. Orland}
\affiliation{Institut de Physique Th\'eorique, CEA, CNRS, UMR3681,
F-91191 Gif-sur-Yvette, France}
\alsoaffiliation{ Beijing Computational Science Research Center, No.10 East
Xibeiwang Road, Beijing 100193, China}
\email{henri.orland@cea.fr}

\author{T. Sakaue}
\affiliation{Department of Physics and Mathematics, Aoyama Gakuin
University, 5-10-1 Fuchinobe, Chuo-ku, Sagamihara, Kanagawa 252-5258,
Japan}
\alsoaffiliation{PRESTO, Japan Science and Technology Agency (JST), 4-1-8
Honcho Kawaguchi, Saitama 332-0012, Japan}
\email{sakaue@phys.aoyama.ac.jp}

\author{C. Vanderzande}
\affiliation{Faculty of Sciences, Hasselt University, 3590 Diepenbeek, Belgium}
\alsoaffiliation{KU Leuven, Institute for Theoretical Physics, Celestijnenlaan
200D, 3001 Leuven, Belgium}
\email{carlo.vanderzande@uhasselt.be}

\title{The effect of memory and active forces on transition path times distributions}

\begin{document}

%
%
%
%
%

%
%
%
%

\begin{abstract}
An analytical expression is derived for the transition path time
distribution for a one-dimensional particle crossing of a parabolic
barrier. Two cases are analyzed: (i) A non-Markovian process described by
a generalized Langevin equation with a power-law memory kernel and (ii)
a Markovian process with a noise violating the fluctuation-dissipation
theorem, modeling the stochastic dynamics generated by active forces. In
the case (i) we show that the anomalous dynamics strongly affecting
the short time behavior of the distributions, but this happens only
for very rare events not influencing the overall statistics. At long
times the decay is always exponential, in disagreement with a recent
study suggesting a stretched exponential decay. In the case (ii) the
active forces do not substantially modify the short time behavior of the
distribution, but lead to an overall decrease of the average transition
path time. These findings offer some novel insights, useful for the
analysis of experiments of transition path times in (bio)molecular
systems.  
\end{abstract}


\maketitle

\section{Introduction}

Biomolecular folding involves structural transitions of various time- and
lengthscales. A simplified description of this process employs a single
reaction coordinate performing stochastic dynamics along a free energy
landscape. In the case of a two state folding, the folded and unfolded
states correspond to two free energy minima, which are separated by a
barrier. Typically, this barrier is high compared to the characteristic
thermal energy $k_BT$, therefore the molecule spends the predominant
fraction of its time close to one of the minima~\cite{hang90}. Transition
paths are the part of the stochastic trajectory corresponding to an
actual barrier crossing event~\cite{humm04}. Although the transition paths
correspond to a tiny fraction of the stochastic trajectory, they encompass
all the information of the folding process. Measuring their duration has
been for long time a big challenge, owing to the high time resolution
needed.  In the past few years, however, experiments have sufficiently
progressed to make measurements of transition path times in nucleic acids
and protein folding possible~\cite{chun09,neup12,true15,neup17}. Recently
also the full probability distribution function of
transition path times, obtained from the statistics of a
large number of events, was determined~\cite{neup16}. The
theory of transition path times have been discussed in several
papers~\cite{bere05,dudk06,zhan07,sega07,chau10,orla11,kim15,maka15,dald16,bere17,lale17}.
These studies mostly employed memoryless Markovian dynamics, while
correlated noise, leading to memory effects and anomalous dynamics,
was only considered in a few recent works~\cite{poll16,sati17}.

Anomalous dynamics is ubiquitous in macromolecular
systems as polymers, as it is known from many
examples~\cite{saka07,panj07,dubb11,walt12,fred14,saka17,vand17}.
This dynamics is characterized by a mean-square displacement of a
suitable reaction coordinate scaling as $\langle \Delta x^2 \rangle
\sim t^\alpha$, with $\alpha \neq 1$. The analysis of the effect of an
underling anomalous dynamics on transition path times is therefore an
interesting case to study, which is one of the aims of this paper. Another
purpose of the present work is to analyze transition path times for
stochastic processes in which the noise has a non-thermal component,
ie not satisfing a fluctuation-dissipation relation. Such noise has been
used in the description of the dynamics of active systems~\cite{bech16}.
Our primary interest is to calculate the transition path time (TPT)
distribution for these two cases and discuss the differences with the more
standard situation of Markovian dynamics in thermal systems. We consider
here a parabolic barrier, which leads to a dynamics described by linear
stochastic differential equations and to Gaussian processes. We show
that, using the formalism developed recently in Ref.~\cite{lale17}, the
calculations are manageable and lead to some simple expressions for the
TPT distributions. We discuss here several features of these distributions
such as the short and long time behavior in the limit of high barriers.

\section{Generalities}

We consider a particle performing a stochastic dynamics on an inverted
parabolic potential barrier $V(x) = -k x^2/2$, with $k>0$. At time
$t=0$ the particle starts from a point $-x_0+\varepsilon$. Transition
paths are those paths reaching $x_0$ at the right side of the barrier
without recrossing $-x_0$ and $x_0$. To compute the distribution of
their duration one should solve the Langevin equation imposing absorbing
boundary conditions in $-x_0$ and $x_0$.  Free boundary conditions are
however easier to handle and provide a good approximation if the barrier
is high~\cite{zhan07}, i.e. $\beta E \gg 1$, with $E=k x_0^2/2$ and
$\beta=1/k_BT$ the inverse temperature.  This is because the probability
of multiple crossings in $\pm x_0$ is negligible for high barriers.

In Ref.~\cite{lale17} the TPT distribution was calculated
for a Markovian particle with inertia. It was shown that both in the inertial 
and overdamped cases the TPT distribution assumes the general form~\cite{lale17}
\begin{equation}
p_{TP}(t) = 
-\frac{2}{\sqrt{\pi}}\frac{\dot G(t)e^{-G^2(t)}}{1-\mathrm{Erf}(\sqrt{\beta E})}.
\label{eq:PTPT_quadr}
\end{equation}
where $\dot G \equiv dG/dt$ and
\begin{equation}
G(t) \equiv \frac{x_0 - \overline{x}(t)}{\sqrt{2 \sigma^2(t)}}.
\label{eq:G}
\end{equation}
In the previous equation
\begin{eqnarray}
\overline{x} (t) &\equiv& \left\langle  x(t)  \right\rangle 
\label{def_mean} \\
\sigma^2 (t) &\equiv& 
\left\langle \left( x(t) - \overline{x}(t)\right)^2 \right\rangle 
\label{def_var}
\end{eqnarray}
are the mean and variance of the process. 

In the overdamped case the function $G(t)$ assumes a simple form
\begin{equation}
G(t) = \sqrt{\beta E} \, 
\sqrt{\frac{e^{\Omega t} + 1}{e^{\Omega t} - 1}}
\label{Gt_overdamped}
\end{equation}
where $\Omega = k/\gamma$ and $\gamma$ is the friction coefficient. The
$G(t)$ in the inertial case is more complex and is given in
Ref.~\cite{lale17}.

For short times $G(t) \approx\sqrt{2} x_0/\sigma(t)$, which diverges
as a consequence of the the initial condition $x(0)=-x_0$, implying
$\sigma(t) \to 0$. This leads to a TPT distribution vanishing with an
essential singularity as $t \to 0$. In the Markovian case the behavior
was found to be different in the overdamped $p_\mathrm{TP}(t) \sim
\exp(-x_0^2/Dt)$ and inertial $p_\mathrm{TP}(t) \sim \exp(-2 \beta
m x_0^2/t^2)$ cases~\cite{lale17} (here $m$ is the particle mass and
$D=1/(\beta \gamma)$ the diffusion coefficient). At long times $G(t)$
converges to a constant in both cases, while its derivative decays
exponentially $\dot G (t) \sim \exp(-\lambda t)$, where $\lambda^{-1}$
is the longest relaxation time of the process ($\lambda=\Omega$
in the overdamped limit~(\ref{Gt_overdamped})). This leads to an
exponential decay $p_\mathrm{TP}(t) \sim \exp(-\lambda t)$ for the long
time behavior of the distribution both in the overdamped and inertial
case~\cite{lale17}. In the next Section we compute $G(t)$ for a process
with correlated and active noise and discuss the TPT distribution obtained
from it.

\section{Memory effects in transition path times}
\label{sec:memory}

A reaction coordinate $x$ is by definition a slow variable for which
standard statistical mechanical arguments show that its time evolution is
given in terms of a generalized Langevin equation~\cite{zwan01}. For a
parabolic barrier in the overdamped case this equation takes the form
\begin{eqnarray}
\int_0^t K(t-\tau)\ \dot{x}(\tau) d\tau = k x(t) + \xi(t)
\label{GLE}
\end{eqnarray}
Here $K(t)$ is a memory kernel. The noise $\xi(t)$ is assumed to be a
Gaussian process with average zero and a correlation that in equilibrium
is related to $K(t)$ by the fluctuation-dissipation theorem
\begin{eqnarray}
\langle \xi(t) \xi(t')\rangle = k_B T K(|t-t'|)
\label{C2}
\end{eqnarray}
We focus here on a power law memory kernel
\begin{eqnarray}
K(t) = \frac{\eta_\alpha\, t^{-\alpha}}{\Gamma(1-\alpha)}
\label{kernel}
\end{eqnarray}
where $0 < \alpha \leq 1$ and where, following Ref.~\cite{vand17},
we define the generalized friction coefficient as $\eta_\alpha=\gamma
\Gamma(3-\alpha)$. In the limit $\alpha \to 1^-$, the $\Gamma$ function in
the denominator becomes singular and $K(t) = 2\gamma \delta(t)$, i.e. one
recovers the Markovian (memoryless) dynamics. Power law kernels are found,
for instance, in the dynamics of polymers which are characterized by a
longest relaxation time $\tau_R$.  While on time scales much larger than
$\tau_R$ the effects of memory on the motion of a reaction coordinate can
be neglected, these are strongly influencing the polymer dynamics for $t
< \tau_R$. Polymers have a memory kernel $K(t)$ that, for $t < \tau_R$,
can be well approximated by a power law~\cite{panj10, saka15}. This
power law behavior is a characteristic of systems with a broad spectrum
of relaxation times.

The generalized Langevin equation (\ref{GLE}) with the kernel
(\ref{kernel}) is a linear equation which can be solved using Laplace
transforms. The initial condition is $x(t=0)=-x_0$. As explained in
the previous section we do not impose specific boundary conditions
in $\pm x_0$, an approximation which is good for steep barriers
$\beta E \gg 1$. The solution of (\ref{GLE}) is (for details see
Appendix~\ref{app:GLE})
\begin{equation}
x(t) = -x_0 \Theta_\alpha(t) + \frac{1}{\eta_\alpha}
\int_0^t \xi(t-\tau) \Psi_\alpha (\tau ) d\tau
\label{6}
\end{equation}
where we introduced the functions
\begin{eqnarray}
\Theta_\alpha (t) &\equiv& 
E_{\alpha,1} \left[( \Omega t)^\alpha \right] 
\label{def:ThetaA}
\\
\Psi_\alpha (t) &\equiv& t^{\alpha -1}  
E_{\alpha,\alpha}  \left[( \Omega t)^\alpha \right] 
\label{def:PsiA}
\end{eqnarray}
($\Omega \equiv (k/\eta_\alpha)^{1/\alpha}$ is the characteristic 
rate of the process) and where
\begin{eqnarray}
E_{\alpha,\beta}(z)
\equiv \sum_{n=0}^\infty \frac{z^n}{\Gamma(\alpha n + \beta)}
\end{eqnarray}
is the Mittag-Leffler function~\cite{haub11}. 

We assume that the noise $\xi(t)$ is Gaussian, and since the Langevin
equation (\ref{GLE}) is linear, we conclude that also $x(t)$ is Gaussian.
Hence Eqs.~(\ref{eq:PTPT_quadr}) and (\ref{eq:G}) apply. One has for
the average
\begin{eqnarray}
\overline{x}(t) \equiv \langle x(t) \rangle = - x_0 \Theta_\alpha(t)
\label{C5}
\end{eqnarray}
while the variance is given by
\begin{eqnarray}
\sigma^2(t) = \frac{k_B T}{k} \left( \Theta_\alpha^2(t) - 1\right)
\label{C6}
\end{eqnarray}
(details of the calculations are in Appendix~\ref{app:GLE}).

Plugging in (\ref{C5}) and (\ref{C6}) in (\ref{eq:G}) we get:
\begin{eqnarray}
G(t) = \sqrt{\beta E} \sqrt{\frac{\Theta_\alpha(t)+1}{\Theta_\alpha(t)-1}}
\label{Gt_GLE}
\end{eqnarray}
This result generalizes the memoryless case (\ref{Gt_overdamped}),
which is recovered in the limit $\alpha=1$ since $\Theta_{1}(x)= 
E_{1,1}(x) = \exp(x)$.

\begin{figure}[t]
\centering
\includegraphics[width=8cm]{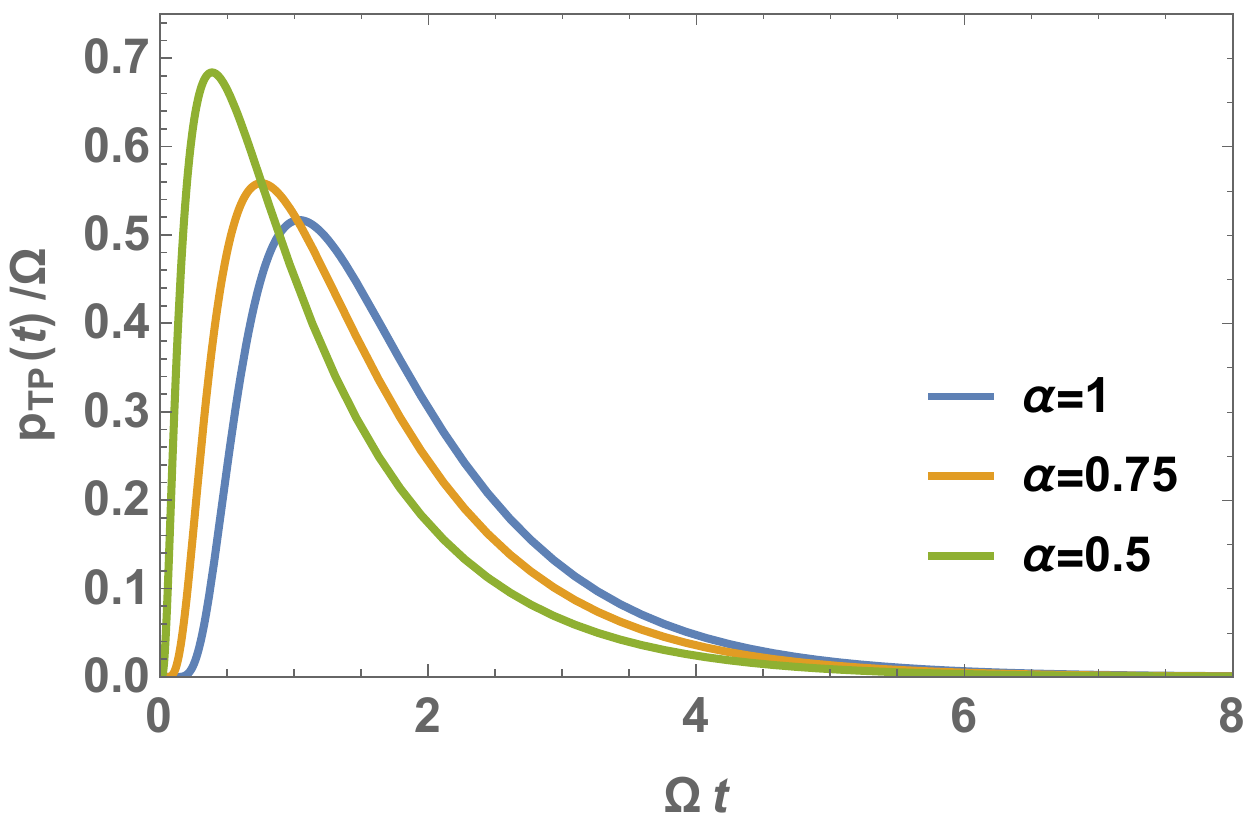}\\
\includegraphics[width=8cm]{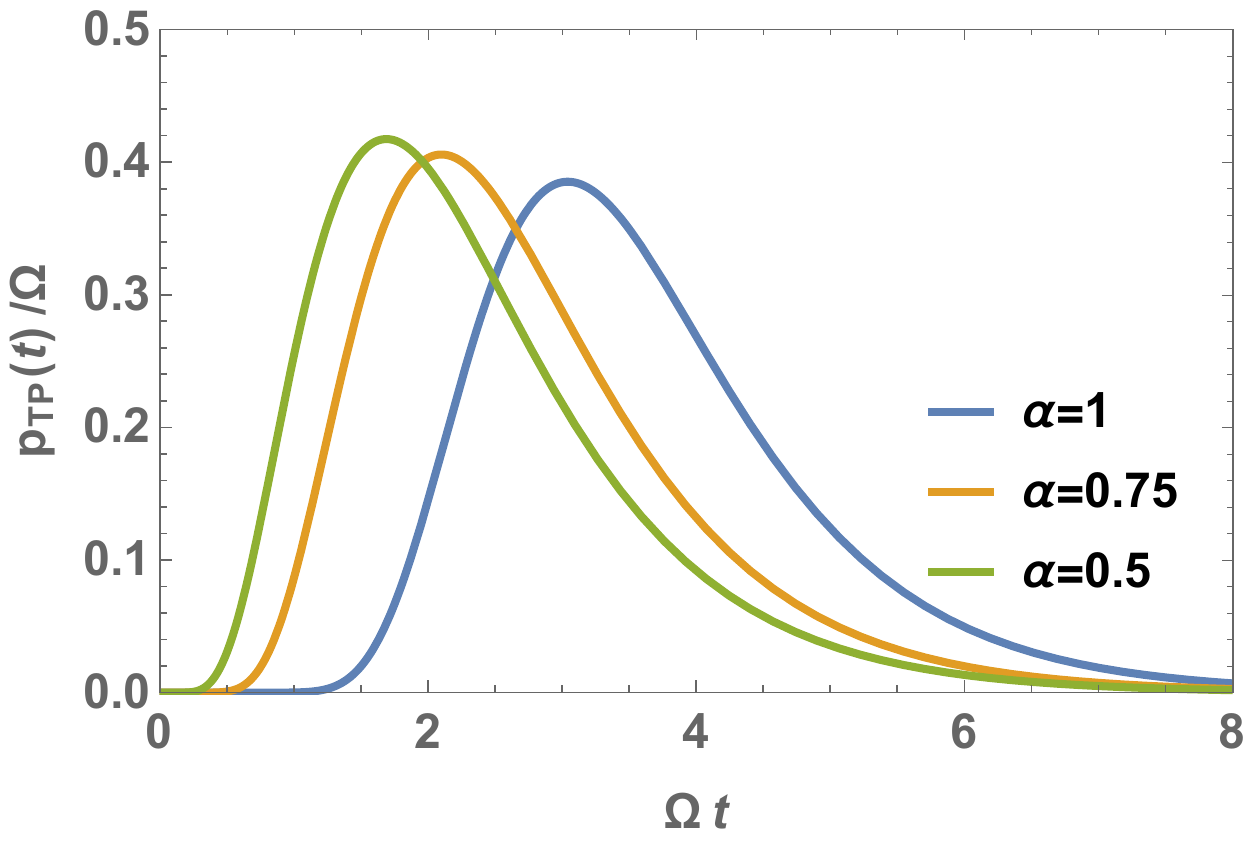}
\caption{Transition path time distribution $p_{\mathrm{TP}}(\Omega
t)/\Omega$ for $\alpha=1$ (blue), $\alpha=.75$ (orange) and $\alpha=0.5$
(green). The other parameters are $k=1$, $\eta_\alpha=10$, $k_B T=1$
and $x_0=\sqrt{2}$ (top), $x_0=\sqrt{20}$ (bottom), corresponding to
$\beta E =1$ and $\beta E =10$, respectively.}
\label{fig1}
\end{figure}

Figure~\ref{fig1} shows plots of the transition path distribution
$p_\mathrm{TP}(t)$ for three different values of $\alpha$ and for two
different values of $k$. The transition path time (in dimensionless
units) decreases with decreasing $\alpha$. We now look at the behavior
of $p_\mathrm{TP}(t)$ for small and large times which can be obtained
from the corresponding behavior of the Mittag-Leffler functions.
As the Mittag-Leffler functions diverge for diverging values of
their arguments (\ref{Gt_GLE}) implies that $G(t) \to \sqrt{\beta
E}$ for $t \to \infty$. For $\dot G(t)$ one finds (for details see
Appendix~\ref{app:GLE}).
\begin{eqnarray}
\dot G(t) \overset{t\sim \infty}{\longrightarrow} 
-\ e^{ - \Omega t}
\label{C8}
\end{eqnarray}
which implies that $p_\mathrm{TP}(t)$ vanishes exponentially, as was
the case for the Markovian model~\cite{lale17}. This is in contrast
with the conclusions of a recent paper~\cite{sati17} where a
stretched exponential decay was found. However, the results of that
paper where obtained from a Fokker-planck equation for systems with
memory that is only correct for a linear potential, or for small
times~\cite{goyc12}. Hence one cannot expect that it gives a correct
asymptotic result.

\begin{figure}[t]
\centering
\includegraphics[width=8cm]{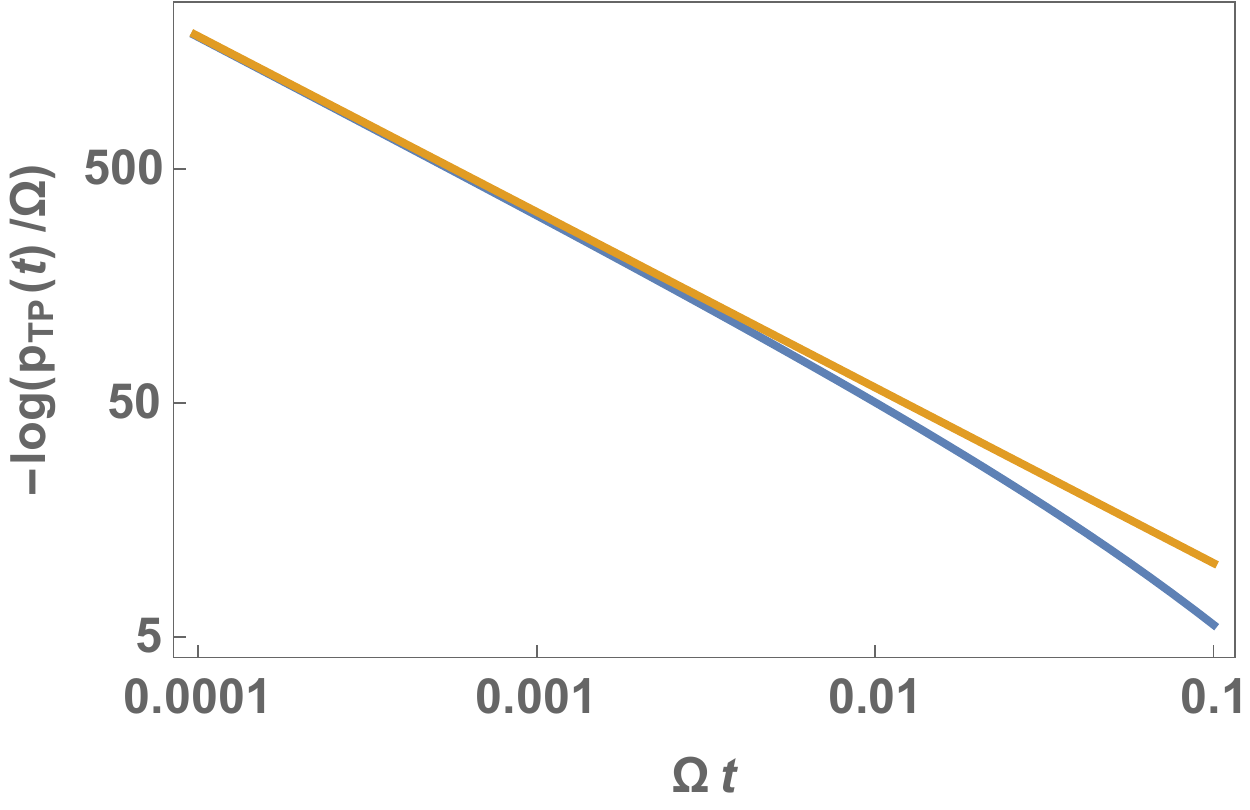}
\caption{Log-log plot of  $-\log{(p_\mathrm{TP} (t})/\Omega)$ versus
$\Omega t$ for $\alpha=0.75, k=2, x_0=1$ in the early time regime $\Omega
t <0.1$. The blue line shows the exact expression while the orange line
gives the early time approximation (\ref{C9}).}
\label{fig2}
\end{figure}

For $t \to 0$, one has that $G^2(t) \to 2 \beta E\ \Gamma(1+\alpha)
(\Omega t)^{-\alpha}$ (see (\ref{app:asymp})) from which it follows
that $\dot G(t) \sim t^{-\alpha/2-1}$. The behavior of the transition
path time distribution for early times is therefore determined by the
essential singularity in $e^{-G^2(t)}$. Hence
\begin{eqnarray}
p_\mathrm{TP} (t) \overset{t\rightarrow 0}{\sim} e^{-G^2(t)}
&=&\exp\left(-\frac{2 \beta E\ \Gamma(1+\alpha)}{(\Omega t)^\alpha}\right)
\nonumber\\
\label{C9}
\end{eqnarray}
We see that the early time behavior does depend on $\alpha$ and that the
exponent could be determined from a straight line fit to a log-log plot
of $-\log{(p_\mathrm{TP} (t})/\Omega)$ versus $\Omega t$. In Fig. \ref{fig2}
we have made such a plot for $\alpha=0.75$ and $\beta E=1$. We find the expected
power law behavior for $\Omega t < 0.005$. An integration of the TPT
distribution shows however that the probability that the transition path
time is in this regime is extremely low ($\sim 10^{-43}$). We therefore
conclude that it is experimentally impossible to determine the exponent
$\alpha$ from the early time behavior of the TPT.

\begin{figure}[t]
\centering
\includegraphics[width=8cm]{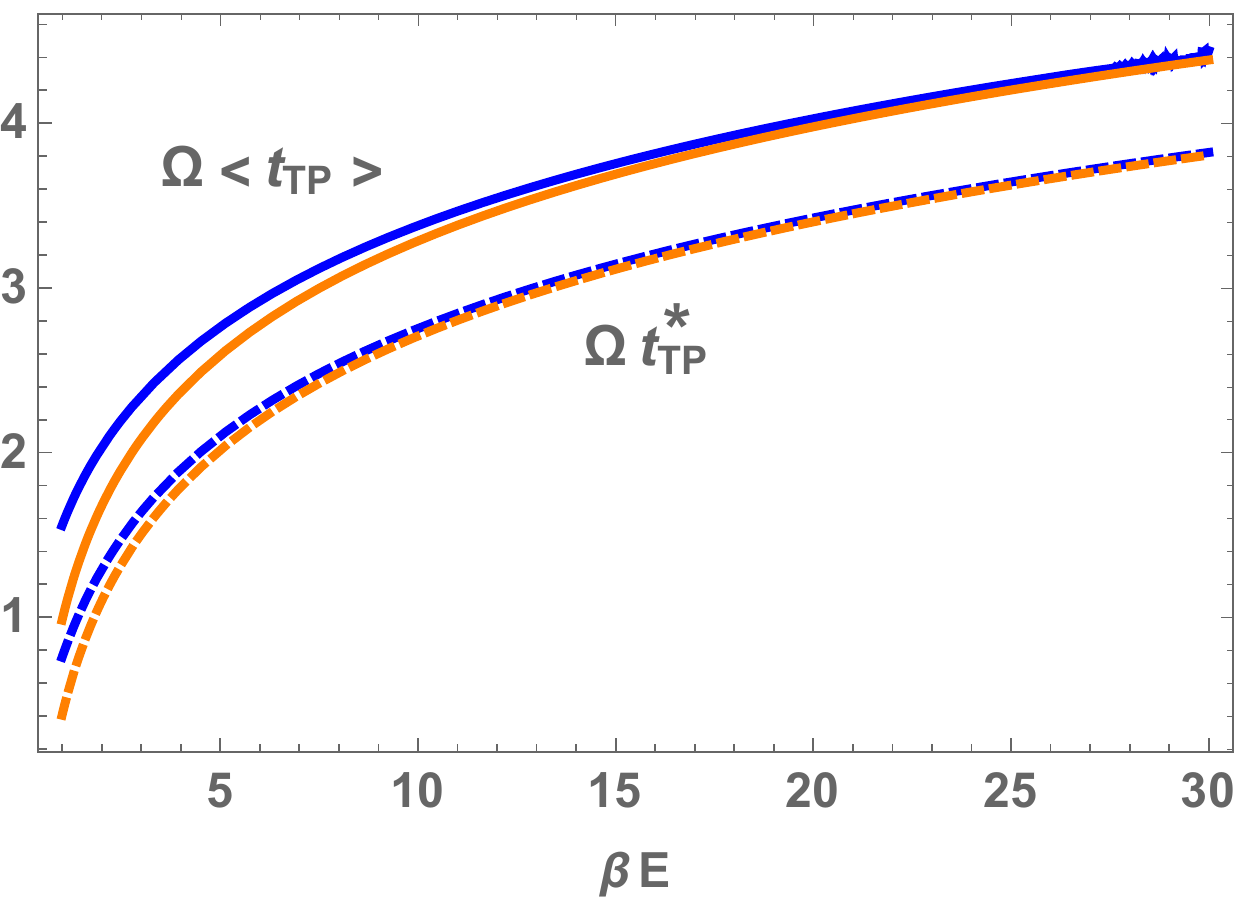}
\caption{Average dimensionless transition path time, $\Omega \langle
t_{TP}\rangle$ (solid lines), and most likely value $\Omega t^*_{TP}$
(dashed lines) as a function of the dimensionless energy $\beta E$ for
$\alpha=0.75$, $k=0.1$, $\eta_\alpha=1$. The blue curves are obtained
numerically from the exact expression for $p_{\mathrm{TP}}(t)$, while the
orange curves are the approximations (\ref{C14}) and (\ref{eq:maxt}).}
\label{fig3}
\end{figure}

We look next at the behavior of the average transition
path time following the same procedure as outlined in
Ref.~\cite{lale17}. As $G$ is a monotonic decreasing function
of $t$ for the calculation it is convenient to perform a change of
variable:
\begin{eqnarray}
\langle t_\mathrm{TP} \rangle = \int_0^\infty t\ p_\mathrm{TP}(t)\ dt= 
\frac{\int_{\sqrt{\beta E}}^\infty t(G) e^{-G^2}\ dG}
{\int_{\sqrt{\beta E}}^\infty e^{-G^2}\ dG}
\nonumber\\
\label{C10}
\end{eqnarray}
The integral in the numerator can not be performed exactly. We can however
get an approximation for $\beta E$ sufficiently large. In that limit the
integrals in (\ref{C10}) are determined by the large $t$-limit of $G(t)$.
The average TPT is then given by (for details, see Appendix~\ref{app:GLE})
\begin{eqnarray}
\langle t_{TP} \rangle = \frac{1}{\Omega} 
\log \left(2 \alpha e^C \beta E\right) + 
{\cal O} \left(\frac{1}{\beta E}\right)
\label{C14}
\end{eqnarray}
where $C \approx 0.577215$ is the Euler-Mascheroni constant (in the
Markovian limit $\alpha=1$ this expression coincides with that previously
obtained by Szabo~\cite{chun12}). In Fig. \ref{fig3}, we have plotted the
result of a numerical calculation of the average transition path time  as
a function of $\beta E$ using the full expression for $G(t)$ and compared
it with the approximation (\ref{C14}) for $\alpha=.75$. We remark that,
according to (\ref{C14}), the dimensionless average transition path time,
$\Omega \langle t_{PT} \rangle$, decreases with decreasing $\alpha$ as
was already evident from the plots in Fig. \ref{fig1}.  The most likely
transition path time $t_\mathrm{TP}^*$, corresponding to the maximum of
a distribution is (see Appendix)
\begin{equation}
t_\mathrm{TP}^* = \frac{1}{\Omega} \log (2 \alpha \beta E) 
+ {\cal O} \left(\frac{1}{\beta E}\right)
\label{eq:maxt}
\end{equation}
and show a similar logarithmic dependence on the barrier height as the
average (\ref{C14}). The comparison between the analytical expression
(\ref{eq:maxt}) and the numerical calculation of the maximum is shown
in Fig.~\ref{fig3} as dashed lines.

\section{Transition path times in the presence of active forces}
\label{sec:active}

The folding of a biopolymer {\it in vivo} takes place in an
environment which is out of equilibrium due to the action of
various ATP-dependent active processes within a cell. These
processes are known to modify the dynamics of various "probes"
like microspheres~\cite{caspi00,gal12,goldstein13} and chromosomal
loci~\cite{weber12,javer13,javer14,zidovska13,saka16,kimura16}. Typically, the
active forces lead to an enhanced diffusion or even superdiffusive
behavior. Similar phenomena have been observed in artificial acto-myosin
networks~\cite{soares14,stuhrmann12,toyota11,brangwynne08,fakhri14}. It
has been found that the effect of the underlying motor processes can often
be described in terms of an active noise $\eta(t)$ which is correlated
over the time scale $\tau_A$ that the motors work. These times are of
the order of seconds.

In a recent study on the behavior of active Brownian particles near
soft walls, the dynamics of a semiflexible polymer immersed in an
environment of such particles was investigated \cite{Nikola16}. Active
Brownian particles move in a direction $\vec{e}$ which is subject to
rotational Brownian diffusion. The force they produce on a (flexible)
soft wall (like, for example a polymer) will therefore be exponentially
correlated, where now the timescale $\tau_A$ of the correlation is
related to the rotational diffusion constant. It was found that due to
pressure instabilities, a sufficiently long polymer folds, even in the
absence of interactions among the monomers.

Inspired by these two examples of folding in a non-equilibrium
environment, it may be of interest to study also the effect of active
forces on transition path times. We start from the Langevin equation
\begin{eqnarray}
\gamma \dot{x} (t) = k\ x(t) + \xi(t) + \eta(t)
\label{C15}
\end{eqnarray}
Here $\xi(t)$ is now a Markovian thermal force  while $\eta(t)$ is the
active noise which assume to have an exponential correlation. 
\begin{eqnarray}
\langle \eta(t)\ \eta(t') \rangle = C \exp\left(-\frac{|t-t'|}{\tau_A}\right)
\label{C16}
\end{eqnarray}
We also take $\langle \eta(t) \rangle=0$. The coefficient $C$ measures
the strength of the active force. There is no associated friction force
so that (\ref{C15}) describes a system out of equilibrium.

The solution of (\ref{C15}) with initial condition $x(t=0)=-x_0$ is
\begin{eqnarray*}
x(t) = - x_0 e^{\Omega t} + \frac{1}{\gamma} 
\int_0^t e^{\Omega (t-t')} \left(\xi(t') + \eta(t')\right)dt'
\end{eqnarray*}
From this we find that the deterministic motion is
\begin{eqnarray}
\overline{x}(t) = \langle x(t) \rangle = - x_0 e^{\Omega t}
\end{eqnarray}
while the variance of the position is given by
\begin{eqnarray}
\sigma^2(t) &=& \langle (x(t) - \overline{x}(t))^2 \rangle 
\nonumber\\
&=& \left(\frac{k_B T}{k} + 
\frac{C \tau_A \Omega}{k^2(1-\tau_A \Omega)} \right) 
\left(e^{2 \Omega t}-1\right)
\nonumber\\
&+& \frac{2 C \tau_A^2 \Omega^2}{k^2 (\tau_A^2 \Omega^2 - 1)}
\left(e^{2 \Omega t}- e^{(1 -1/\Omega \tau_A) \Omega t}\right)
\nonumber\\
\end{eqnarray}
It is convenient to describe the escape over the parabolic in terms of an
effective temperature as was done in a study of the motion of colloids
in active bath of bacteria and in the presence of a confining harmonic
potential~\cite{maggi14}.

Asymptotically $\sigma^2(t) e^{-2 \Omega t}$ goes to a constant which
can be used to define this effective temperature $T^\star$
\begin{eqnarray*}
\lim_{t \to \infty} \sigma^2(t) e^{-2 \Omega t} &=& 
\frac{1}{k} \left[ k_B T + \frac{C \tau_A \Omega}
{k( \tau_A \Omega + 1)} \right] \\ & \equiv & \frac{k_B T^\star}{k}
\end{eqnarray*}
The effective temperature takes over the role of the physical temperature
in the transition path time distribution. Going through the calculations
of Ref.~\cite{lale17} we find that in this case
\begin{eqnarray}
p_\mathrm{TP}(t) = - \frac{2}{\sqrt{\pi}} 
\frac{\dot G(t) e^{-G^2(t)}}{1- \mathrm{Erf}{(\sqrt{\beta^\star E})}}
\end{eqnarray}
where $G(t)$ is given by (\ref{eq:G}) and
$\beta^\star=1/k_B T^\star$.

\begin{figure}[t]
\centering
\includegraphics[width=8cm]{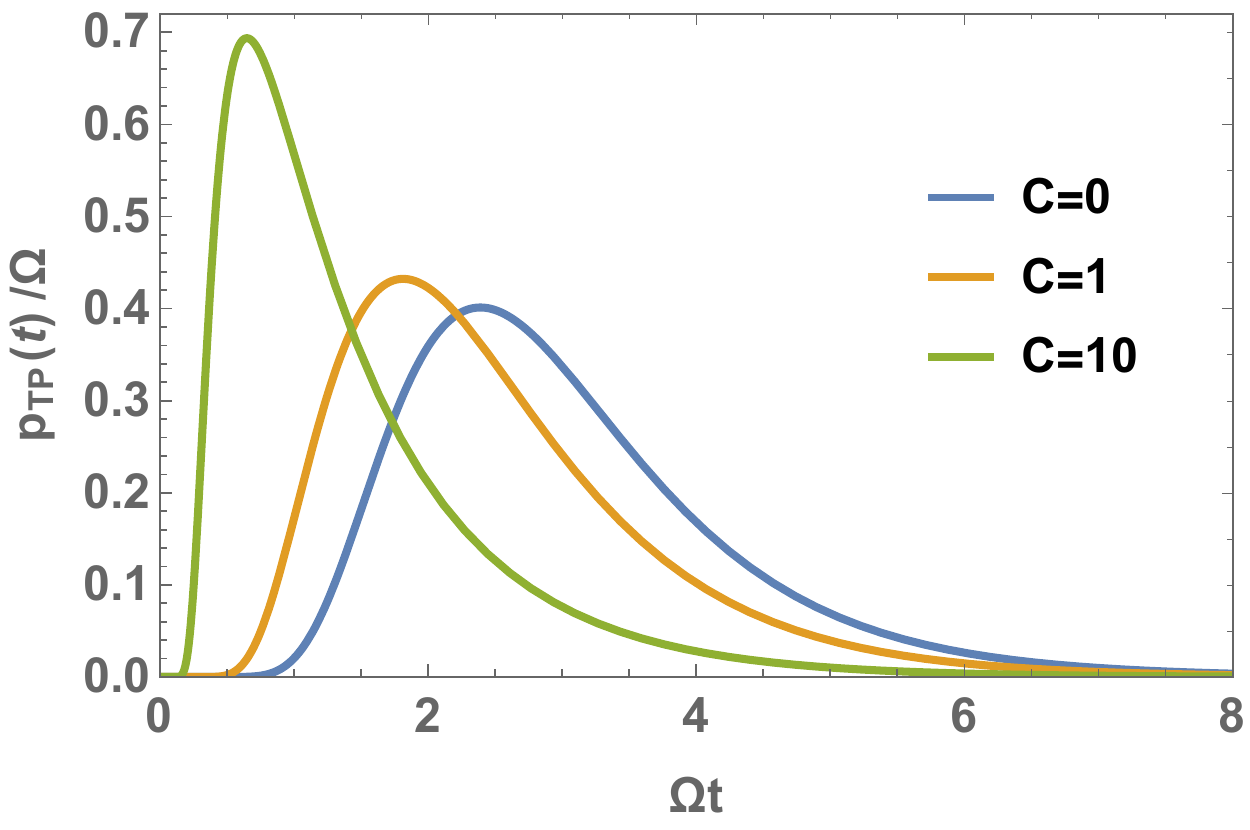}
\caption{Transition path time distribution $p_\mathrm{TP}(\Omega t)/\Omega$
for $C=0$ (blue), $C=1$ (orange) and $C=10$ (green) and for $k=0.1,
\eta_\alpha=1, x_0=10, k_BT=1, \tau_A=1$.}
\label{fig5}
\end{figure}

From these results one can find that at early times, the transition path
time is again governed by the essential singularity in $e^{-G^2(t)}$ whose
form is not modified by the active forces. The late time decay is governed
by $\dot G(t)$ which decays exponentially. The only change is in the
prefactor of the exponential which now involves the effective temperature
\begin{eqnarray}
\dot G(t) \sim - \sqrt{\beta^\star E}\ 
\Omega \ e^{-\Omega t}  \hspace{1cm} (t \to \infty)
\end{eqnarray}
Finally, in the expression for the average transition path time, the
effective temperature also replaces the real temperature
\begin{eqnarray}
\langle t_{TP}(t)\rangle = \Omega^{-1} \log\left( 2 e^C \beta^\star E\right)
\end{eqnarray}
Since $\beta^\star < \beta$, the addition of active forces leads to
a decrease of the average transition path time.  In Fig. \ref{fig5},
we have plotted some distributions where it can be seen that indeed
the average transition path time decreases if the effective temperature
(here tuned by changing $C$ at fixed $\tau_A$) increases.

\begin{figure}[h]
\centering
\includegraphics[width=8cm]{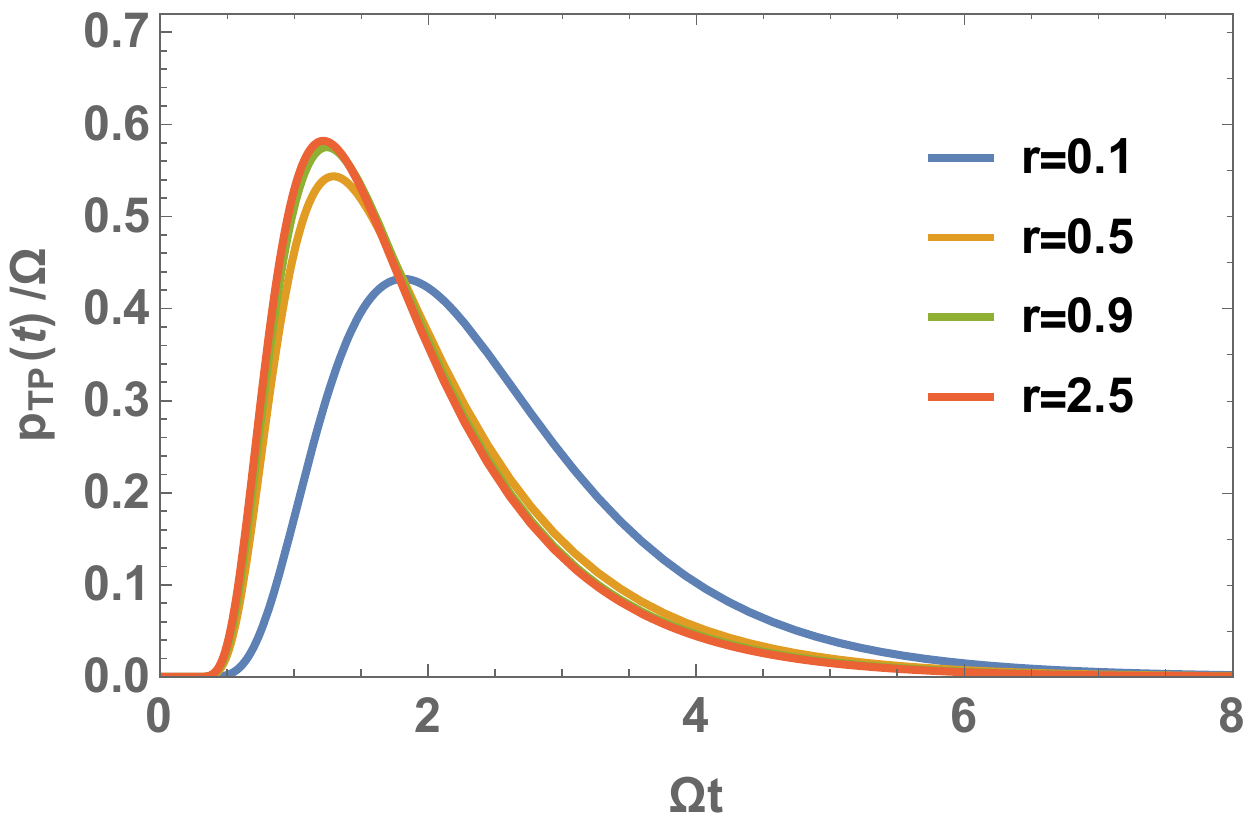}
\caption{Transition path time distribution $p_{TP}(\Omega t)/\Omega$ for
$r=0.1$ (blue), $r=0.5$ (orange) and $r=0.9$ (green) and $r=2.5$(red)
and for $k=0.1, \eta_\alpha=1, x_0=10, k_BT=1, C=1$. Here $r=\Omega \tau_A$
is the ratio of the two timescales in the problem.}
\label{fig6}
\end{figure}

As can be seen from the expression of the effective temperature,
the dependence on $\tau_A$ is weak once it becomes bigger then
$\Omega^{-1}$ (the other time scale in problem). This can also be seen
in Fig. \ref{fig6} where $C$ is fixed and $\tau_A$ is increased from
values below to values above $\Omega^{-1}$ .

\section{Discussion}
\label{sec:discussion}

Conformational transitions of molecular systems between two different
states are governed by two time scales. The Kramers time corresponds
to the typical time spent in a given conformation (the dwell time),
while the transition path time characterizes the actual duration
of the transition. Transition path times, which have been measured
in proteins and nucleic acids folding experiments during the past
decade~\cite{chun09,chun12,neup12,true15,neup17}, can be a few orders
of magnitudes shorter than Kramers' times.

In this paper we have analyzed the TPT distribution of a 
one dimensional stochastic particle undergoing Langevin dynamics and
crossing a parabolic barrier. We investigated the effects of memory and
non-equilibrium thus extending previous analysis~\cite{zhan07,lale17}. As
the barrier is parabolic, the associated Langevin equation is linear
and hence exactly solvable in the case of free boundary conditions.
This solution is expected to approximate very well the absorbing boundary
case for steep barriers.

In Ref.~\cite{poll16} the effect of memory on transition path times
was also investigated. The TPT-distribution was derived for an
arbitrary memory kernel starting from an Hamiltonian formulation
in which the particle dynamics is coupled to a bath of harmonic
oscillators~\cite{poll16}. We expect that the expressions reported
in \cite{poll16} will agree with our results in the case of overdamped
dynamics with power-law memory in the limit of high barriers.  Our result
for the power law kernel has the advantage that it is simple and of the
same form as in the Markovian case. We expect it to be easier to compare
with experiments.

\subsection{Long time limit of transition path time
distribution: why exponential decay?}

We have found that the asymptotic decay of the transition path time
distribution remains exponential for both cases investigated and
therefore has a remarkable universal behavior. This contrasts with
the conclusions of Ref.~\cite{sati17}. In that work, which employed a
Fokker-Planck equation with a time dependent diffusion constant $D(t)
\sim t^{\alpha-1}$, it was argued that the large time decay of the TPT
is stretched exponential.  However, it was shown that for a particle
in a harmonic potential the correct expression for $D(t)$ coincides
with that used in~\cite{sati17} only for short times~\cite{goyc12}.
This suggests that the asymptotic stretched exponential behavior
reported in \cite{sati17} cannot be trusted. A diffusion constant $D(t)
\sim t^{\alpha-1}$ was also derived for a particle under constant
force~\cite{goyc12}.

To get some more insights about the differences in the effect of memory
kernels in the
constant force and the parabolic barrier case let us consider
the following equation
\begin{eqnarray}
\int_{0}^{t}K(t-\tau) \, {\dot x}(\tau) d\tau = f
\label{GL_f}
\end{eqnarray}
which describes the average motion of particle driven by a constant force
$f$ in a medium characterized by the exponent $\alpha$.  Using Laplace
transforms we find $x(t)-x(0) \sim t^\alpha$, with the normal drift
$x(t)-x(0) \sim t$ recovered in the Markovian limit $\alpha \to 1$.
This behavior can be deduced from an effective medium description
\begin{eqnarray}
\Gamma_\mathrm{eff}(t) {\dot x}(t) \sim f
\label{gamma_eff_f}
\end{eqnarray}
where the effective friction $\Gamma_\mathrm{eff}(t) \sim t^{-\alpha+1}$,
as expected from the time integral of memory kernel, grows with time due
to memory effects (the result is consistent with the Einstein relation
for the diffusion constant discussed in~\cite{goyc12}). This indicates
the velocity ${\dot x}(t) \sim f t^{\alpha-1}$ decreases with time,
hence, the anomalous drift. In the parabolic barrier there is, however,
a crucial difference. The effective medium description would give
\begin{eqnarray}
\Gamma_\mathrm{eff}(t) {\dot x}(t) \sim k x(t)
\label{gamma_eff_1}
\end{eqnarray}
with solution
\begin{eqnarray}
x(t) = -x_0 \ \exp{\left[(\Omega t)^{\alpha} \right]}
\end{eqnarray}
which is a stretched exponential behavior. This is not consistent with
the exact solution of the generalized Langevin equation discussed in this
paper, which yields for the average position an exponentially growing
function at long times (obtained from the asymptotic behavior of the
Mittag-Leffler function of~(\ref{C5})).

To understand this apparent paradox, we point out that the
effective friction argument would be valid for a process in which
the velocity ${\dot x}(t)$ is a slowly varying function. For a self-similar
process, where the velocity changes according to a power law ${\dot x}(t)
\sim t^{-\gamma}$, the coarse grained variable $\int_0^t dt' {\dot x}(t')
/ t \sim t^{-\gamma}$ by time average behaves similarly with the original
variable. In such a case, the effective friction argument should work
to get the correct scaling behavior. 


However, in the parabolic barrier crossing, the velocity increases rapidly
(exponentially), therefore, the contribution from the memory kernel integral
is dominated by the most recent term only. This implies that in the
long time scale, we should expect an effective description, in which
the system feels only the instantaneous response, hence our
effective equation is
\begin{eqnarray}
{\tilde \gamma} {\dot x}(t) = k x(t)
\end{eqnarray}
where $\tilde{\gamma}$ is a renormalized friction coefficient.
Our argument suggests that it is this renormalization that is behind the
universal exponential decay in the long time limit of transition path
time distribution. This asymptotic behavior sets already in at $\Omega
t \simeq 1.5$ as seen numerically (Fig.~\ref{fig1}).

\subsection{Comparison with experiments: possible implications}

Differently from the Kramers' time, which is characterised by an
exponential dependence on the barrier height $E$, the average TPT in
the overdamped limit scales logarithmically $\langle t_\mathrm{TP} \rangle
\sim \log (\beta E)$, where $\beta$ is the inverse temperature.  We have
shown here that the logarithmic dependence also holds in the presence
of memory effects or of active forces. The effect of the active forces
is simply to increase the temperature to a higher effective one. Hence
the average TPT will always decrease in this case.  Memory also decrease
the average TPT when measured in dimensionless units.

We believe that our results are helpful in interpreting experimental
results. Indeed the barrier height as determined from a comparison between
experiments and a model for diffusion in a parabolic potential without
memory terms gave values that were much lower than those determined
by other means. Our calculations have shown that both memory effects
and non-equilibrium effects can have the same effect as lowering the
potential barrier. While in current experiments non-equilibrium effects
almost surely play no role they could certainly be of relevance inside
the cellular environment.  The experimental results on DNA-hairpins and
proteins  can however be understood from a model with memory. Indeed,
in that case we predict that there are more short transit times than in
a model without memory (when time is measured in rescaled unit $\Omega
t$). This is indeed what is found experimentally. Further research
will have to show whether this qualitative agreement can be made more
quantitative.


\begin{acknowledgement}

Discussions with M. Caraglio and M. Laleman are gratefully acknowledged.

\end{acknowledgement}

%
%
%

\section{Appendix A: Power law memory kernel}
\label{app:GLE}

The solution of the generalized Langevin equation~(\ref{GLE}) with
power law memory kernel is obtained by performing its Laplace transform
\begin{equation}
\widetilde{K} (s)\ (s \widetilde{x} (s) + x_0) = 
k \widetilde{x}(s) + \widetilde{\xi}(s)
\label{GLE_LT}
\end{equation}
where $\widetilde{f} (s)$ indicates the Laplace transform of the function
$f(t)$. To obtain the previous equation we used the convolution theorem
(the Laplace transform of a convolution product is the product of
the Laplace transforms) and the fact that the Laplace transform of a
derivative is
\begin{equation}
\widetilde{\dot{f}} = s \widetilde{f} (s) - f(0)
\end{equation}
(in our case the initial condition is $x(0)=-x_0$).
Solving~(\ref{GLE_LT}) we get
\begin{equation}
\widetilde{x} (s) = \frac{-x_0 \widetilde{K} (s)}{s \widetilde{K} (s) - k} +
\frac{\widetilde{\xi}(s)}{s \widetilde{K} (s) - k}
\label{sol_LTx}
\end{equation}
The Laplace transform of the power law kernel~(\ref{kernel}) is
\begin{equation}
\widetilde{K} (s) = \gamma \, s^{\alpha-1}
\end{equation}
therefore Eq.~(\ref{sol_LTx}) takes the form
\begin{equation}
\widetilde{x} (s) = \frac{-x_0 s^\alpha}{s \left(s^\alpha - k/\gamma\right)} +
\frac{1}{\gamma} \, 
\frac{\widetilde{\xi}(s)}{s^\alpha - k/\gamma}
\label{sol_LTx2}
\end{equation}
To perform the inverse transform we use the following relation
\begin{equation}
\int_0^\infty dt\, e^{-ts} \, t^{\beta-1} E_{\alpha,\beta}(at^\alpha) = 
\frac{s^{\alpha}}{s^{\beta} (s^{\alpha}-a)}
\label{def_ML}
\end{equation}
where $E_{\alpha,\beta}(z)$ is known as Mittag-Leffler
function~\cite{haub11}. To handle the two terms in the left hand
side of (\ref{sol_LTx2}) one can use (\ref{def_ML}) with $\beta=1$
and $\beta=\alpha$.  For this purpose it is convenient to introduce
the functions
\begin{eqnarray}
\Theta_\alpha (t) &\equiv& 
E_{\alpha,1} \left[( \Omega t)^\alpha \right] 
\label{def:ThetaA}
\\
\Psi_\alpha (t) &\equiv& t^{\alpha -1}  
E_{\alpha,\alpha}  \left[( \Omega t)^\alpha \right] 
\label{def:PsiA}
\end{eqnarray}
where $\Omega \equiv (k/\gamma)^{1/\alpha}$ is a characteristic rate of
the process. Inverting (\ref{sol_LTx2}) we get
\begin{equation}
x(t) = -x_0 \Theta_\alpha(t) + \frac{1}{\gamma}
\int_0^t \xi(t-\tau) \Psi_\alpha (\tau ) d\tau
\label{6}
\end{equation}
Averaging over noise we get the average position, or equivalently the 
deterministic solution
\begin{equation}
\bar{x}(t) = -x_0 \Theta_\alpha(t) 
\label{7}
\end{equation}
while the variance (\ref{def_var}) is
\begin{eqnarray}
\sigma^2(t) &=& \frac{k_B T}{\gamma} 
\int_0^t d\tau d\sigma \frac{|\tau-\sigma|^{-\alpha}}{\Gamma(1-\alpha)}\ 
\Psi_\alpha (\tau) \Psi_\alpha (\sigma) \nonumber\\
&=& \frac{k_BT}{k} \left( \Theta_\alpha^2 (t) - 1 \right)
\label{app:sigma2}
\end{eqnarray}
(the details of the calculation of this
integral are given in Appendix~\ref{app:integral}).
We finally combine the above results to find
$G(t)$, see (\ref{eq:G})
\begin{equation}
G(t) =
\frac{x_0 - \bar{x}(t)}{\sqrt{2 \sigma^2(t)}} = \sqrt{\beta E} 
\sqrt{\frac{\Theta_\alpha(t) + 1}{\Theta_\alpha (t) -1}}
\label{app:GGLE}
\end{equation}
where $E=kx_0^2/2$ is the barrier height. This proves Eq.~(\ref{Gt_GLE})
of the main text.

The Mittag-Leffler function behaves asymptotically as~\cite{haub11}
\begin{equation}
E_{\alpha,\beta}(z) \overset{z\rightarrow\infty}{\longrightarrow}
\frac{1}{\alpha} z^{(1-\beta)/\alpha} \exp(z^{1/\alpha})
\label{ML_large}
\end{equation}
which implies 
\begin{equation}
\Theta_\alpha (t)  \overset{t\rightarrow\infty}{\longrightarrow}
\frac{1}{\alpha} \exp(\Omega t)
\label{A14}
\end{equation}
Hence 
\begin{eqnarray}
G (t)  &\overset{t\rightarrow\infty}{\longrightarrow}& \sqrt{\beta E} 
\label{A15}\\
\dot{G}(t)  &\overset{t\rightarrow\infty}{\longrightarrow}& -
\sqrt{\beta E} \ \alpha \Omega \ \exp(-\Omega t)
\label{A16}
\end{eqnarray}

For small arguments, the Mittag-Leffler function behaves as~\cite{haub11}
\begin{equation}
E_{\alpha,\beta} (z)= 1 + \frac{z^\alpha}{\Gamma(\alpha+\beta)} + \cdots
\end{equation}
hence
\begin{equation}
\Theta_\alpha(t)= 1 + 
\frac{\left( \Omega t\right)^\alpha}{\Gamma(1+\alpha)} + \cdots
\end{equation}
This implies that $G(t)$ diverges for small $t$
\begin{equation}
G (t) \sim (\Omega t)^{-\alpha/2}
\label{app:asymp}
\end{equation}

\subsection{The average transition path time}

To calculate the average TPT
we follow the calculation outlined in \cite{lale17}:
\begin{equation}
\langle t_{TP} \rangle = \int_0^\infty t\ p_{TP}(t)\ dt
= \frac{\int_{\sqrt{\beta E}}^\infty t(G) 
e^{-G^2}\ dG}{\int_{\sqrt{\beta E}}^\infty e^{-G^2}\ dG}
\label{18}
\end{equation}
where we have made the change of variables $G'dt=dG$ and used the
definition of the error function. The integral in the numerator can not
be performed exactly. We can get an approximation for $\beta E  \gg 1$
where the integrals are determined by the large $t$-limit of $G(t)$. 
From (\ref{def:ThetaA}), (\ref{app:GGLE}) and (\ref{ML_large}) one
gets for $t$ large
\begin{eqnarray}
G(t)=\sqrt{\beta E}
\left[ 1 + \alpha e^{-\Omega t} + \ldots \right]
\end{eqnarray}
which can be inverted to
\begin{eqnarray}
t= \frac{1}{\Omega} 
\left( \log \alpha - \log\left(\frac{G}{\sqrt{\beta E}}-1\right)\right)
\label{20}
\end{eqnarray}
If we insert (\ref{20}) into (\ref{18}) and make also here an expansion
for large $\beta E$ we finally get
\begin{eqnarray}
\langle t_{TP} \rangle = \frac{1}{\Omega} 
\log \left(2 \alpha e^C \beta E\right) + O \left(\frac{1}{\beta E}\right)
\label{21}
\end{eqnarray}
where $C \approx 0.577215$ is the Euler-Mascheroni constant.

\subsection{The most likely transition path time}

Another interesting quantity we can infer from the results is 
$t_\mathrm{TP}^*$,
the most likely value of the TPT. This is obtained by solving the
equation
\begin{equation}
\frac{d p_\mathrm{TP}}{dt} = 0
\end{equation}
which from (\ref{eq:PTPT_quadr}) implies $\ddot G = 2 G \dot G^2$ (where
the dot indicates the time derivative), or using (\ref{Gt_GLE}):
\begin{equation}
\dot\Theta_\alpha^2 - \ddot\Theta_\alpha \left( \Theta_\alpha^2 - 1 \right) 
+ 2 \, \Theta_\alpha \dot\Theta_\alpha^2 = 
2 \beta E \, \dot\Theta_\alpha^2 \, 
\frac{\Theta_\alpha+1}{\Theta_\alpha-1}
\label{A25}
\end{equation}
Using the asymptotic $t \to + \infty$ expansion (\ref{A14}) one has
\begin{equation}
\dot\Theta_\alpha (t) \sim \frac{\Omega}{\alpha} \exp(\Omega t),
\quad \quad
\ddot\Theta_\alpha(t) \sim \frac{\Omega^2}{\alpha} \exp(\Omega t)
\end{equation}
and to leading order in $\beta E$ the solution of (\ref{A25}) becomes
\begin{equation}
t_\mathrm{TP}^* = \frac{1}{\Omega} \log (2 \alpha \beta E) 
\label{A27}
\end{equation}

\section{Appendix B: Integral (\ref{app:sigma2})}
\label{app:integral}

To compute the integral (\ref{app:sigma2}) we start from the double
Laplace transform of the function $\Theta_\alpha(|\tau-\sigma|)$.
We have
\begin{eqnarray}
f(s,s') \equiv \int_0^{+\infty} \int_0^{+\infty} \!\! 
d\tau d\sigma \, 
e^{-s\tau -s' \sigma} \Theta_\alpha(|\tau-\sigma|)
\nonumber\\
\end{eqnarray}
To get rid of the absolute value we split the integral in two 
domains so to obtain
\begin{eqnarray}
&&f(s,s') = \int_0^{+\infty} \!\!\!\! 
d\tau \int_\tau^{+\infty} \!\!\!\! d\sigma \, 
e^{-s\tau -s' \sigma} \Theta_\alpha(\sigma-\tau) 
\nonumber \\
&& +\int_0^{+\infty} d\sigma \int_\tau^{+\infty} d\tau \, 
e^{-s\tau -s' \sigma} \Theta_\alpha(\tau-\sigma)
\nonumber \\
&& = \frac{1}{s+s'} \left[ 
\frac{1}{s' \left( 1  - (\Omega/s')^\alpha\right)} + 
\frac{1}{s \left( 1  - (\Omega/s)^\alpha\right)} 
\right]
\nonumber \\
\end{eqnarray}
The integrals can be easily computed using a change of variables
and the property (\ref{def_ML}). The above expression can be
rearranged as follows
\begin{eqnarray}
&&f(s,s') = \frac{1}{ss'\left(1 -(\Omega/s')^\alpha\right) 
\left(1 -(\Omega/s)^\alpha\right)} -
\nonumber \\
&&
\frac{\Omega^{\alpha}}{(ss')^\alpha 
\left(1 -(\Omega/s')^\alpha\right) \left(1 -(\Omega/s)^\alpha\right)}
\frac{s^{\alpha-1} + s'^{\alpha-1}}{s+s'}
\nonumber \\
\label{B3}
\end{eqnarray}
The double inverse Laplace transform of the first term is easy as this
term is the product of a function of $s$ and a function of $s'$. One
has two independent inverse Laplace transform and from (\ref{def_ML})
one sees that this generates $\Theta_\alpha (\tau)\Theta_\alpha (\sigma)$.

The second term in (\ref{B3}) is  a product of two fractions. In the
first one, one recognises the double Laplace transform of $\Omega^\alpha
\Psi_\alpha(\tau) \Psi_\alpha(\sigma)$. For the second one we use
\begin{eqnarray}
\int_0^{+\infty}  d\tau d\sigma \, 
\frac{e^{-\sigma s -\tau s'}|\tau - \sigma|^{-\alpha}}{\Gamma(1-\alpha)} &=&  
\frac{s^{\alpha-1} + s'^{\alpha-1}}{s+s'}
\nonumber\\
\end{eqnarray}
Invoking the convolution theorem of double Laplace transforms,  the
second term of (\ref{B3}) is therefore the double Laplace transform of
the convolution
\begin{eqnarray}
\frac{\Omega^\alpha}{\Gamma(1-\alpha)} \int_0^t \int_0^{t'} 
d\tau d\sigma |\tau-\sigma]^{-\alpha} \Psi_\alpha(\tau) \Psi_\alpha(\sigma)
\nonumber\\
\label{B4}
\end{eqnarray}
Putting everything together we have
\begin{eqnarray}
\frac{\Omega^\alpha}{\Gamma(1-\alpha)}  \int_0^t \int_0^{t'} 
d\tau d\sigma |\tau-\sigma]^{-\alpha} \Psi_\alpha(\tau) \Psi_\alpha(\sigma)
&&\nonumber \\
=\Theta_\alpha (t)\Theta_\alpha (t')-\Theta_\alpha(|t-t'|)
&&
\nonumber\\
\label{B5}
\end{eqnarray}
from which (\ref{app:sigma2}) follows by putting $t=t'$. 


\providecommand{\latin}[1]{#1}
\providecommand*\mcitethebibliography{\thebibliography}
\csname @ifundefined\endcsname{endmcitethebibliography}
  {\let\endmcitethebibliography\endthebibliography}{}

\end{document}